\documentclass[apj,iop]{emulateapj}  
\usepackage{amsmath}
\usepackage{graphicx} 
\usepackage{apjfonts} 
\usepackage{epstopdf}

\shorttitle{Lateral downflows in sunspot penumbral filaments and their temporal evolution}
\shortauthors{Esteban Pozuelo et al.}

\begin{document}

\title{Lateral downflows in sunspot penumbral filaments and their temporal evolution}
\author{S.~Esteban Pozuelo$^1$}
\author{L.~R.~Bellot Rubio$^1$}
\author{J.~de la Cruz Rodr\'iguez$^2$}

\affil{$^1$ Instituto de Astrof\'{\i}sica de Andaluc\'{\i}a (CSIC), 
Apdo.\ 3004, E-18008 Granada, Spain; sesteban@iaa.es}
\affil{$^2$ Institute for Solar Physics, Dept.\ of Astronomy, Stockholm 
University, Albanova University Center, SE-10691 Stockholm, Sweden}

\begin{abstract} 
  We study the temporal evolution of downflows observed at the lateral
  edges of penumbral filaments in a sunspot located very close to the
  disk center.  Our analysis is based on a sequence of nearly
  diffraction-limited scans of the \ion{Fe}{1}
  617.3~nm line taken with the CRisp Imaging
    Spectro-Polarimeter at the Swedish 1 m Solar Telescope.
  We compute Dopplergrams from the observed intensity profiles using
  line bisectors and filter the resulting velocity maps for subsonic
  oscillations. Lateral downflows appear everywhere in the center-side
  penumbra as small, weak patches of redshifts next to or along the
  edges of blueshifted flow channels.  These patches have an
  intermittent life and undergo mergings and fragmentations quite
  frequently. The lateral downflows move together with the hosting
  filaments and react to their shape variations, very much resembling
  the evolution of granular convection in the quiet Sun. There is a
  good relation between brightness and velocity in the
  center-side penumbra, with downflows being darker than upflows on
  average, which is again reminiscent of quiet Sun
  convection.  These results point to the existence of overturning
  convection in sunspot penumbrae, with elongated cells
  forming filaments where the flow is upward but very
  inclined, and weak lateral downward flows.  In general, the circular
  polarization profiles emerging from the lateral downflows do not
  show sign reversals, although sometimes we detect three-lobed
  profiles which are suggestive of opposite magnetic polarities in the
  pixel.
\end{abstract}
\keywords{Sun: convection - photosphere - sunspots - penumbra}

\section{Introduction}

The penumbra of sunspots is a magnetized medium where convection is
expected to be strongly suppressed \citep{1941VAG....76..194B,
  1953sun..book..532C}.  However, we observe it as a relatively bright
structure formed by elongated filaments with typical widths of
$0\farcs2$ \citep[e.g.,][]{1961ApJ...134..275D, 2001A&A...374L..21S}.
How energy is transported in the penumbra remains a controversial
issue.  There is consensus that the mechanism responsible for the
penumbral brightness involves magnetoconvection, but the details are
still poorly understood.

The most conspicuous dynamic phenomenon in sunspots is the Evershed
flow \citep{1909MNRAS..69..454E}---a radial outflow of gas with speeds
of several km~s$^{-1}$ and large inclinations to the vertical. The
Evershed flow is closely related to the filamentary structure of the
penumbra \citep[for a review, see][]{2011LRSP....8....4B}.  Given its
ubiquity and physical properties, it is believed to play a central
role in the energy transport of sunspots.

The origin of the Evershed flow has been the subject of intense debate
for decades. Some theoretical models explain it as a siphon flow
driven by a gas pressure difference between the footpoints of arched
flux tubes \citep{1968AJS....73Q..71M, 1992sto..work.....T,
  1997Natur.390..485M, 2006A&A...452.1089T}. In other models, the
Evershed flow is a hot gas confined to magnetic flux tubes that rises
to the solar surface by some sort of convection
\citep[e.g.,][]{1993A&A...275..283S, 1994A&A...290..295J,
  1998A&A...337..897S, 2003A&A...411..257S}. At the surface the hot
  gas cools down by radiative losses, heats the surroundings, and
  sinks again \citep{2002AN....323..303S}. This model produces a
  bright penumbra \citep{2008A&A...488..749R} and is compatible with
  observations of both penumbral flows and magnetic fields 
\citep{2011LRSP....8....4B}.

\begin{figure*}[t]
\begin{center}
\includegraphics[keepaspectratio = true, width =
        0.75\textwidth]{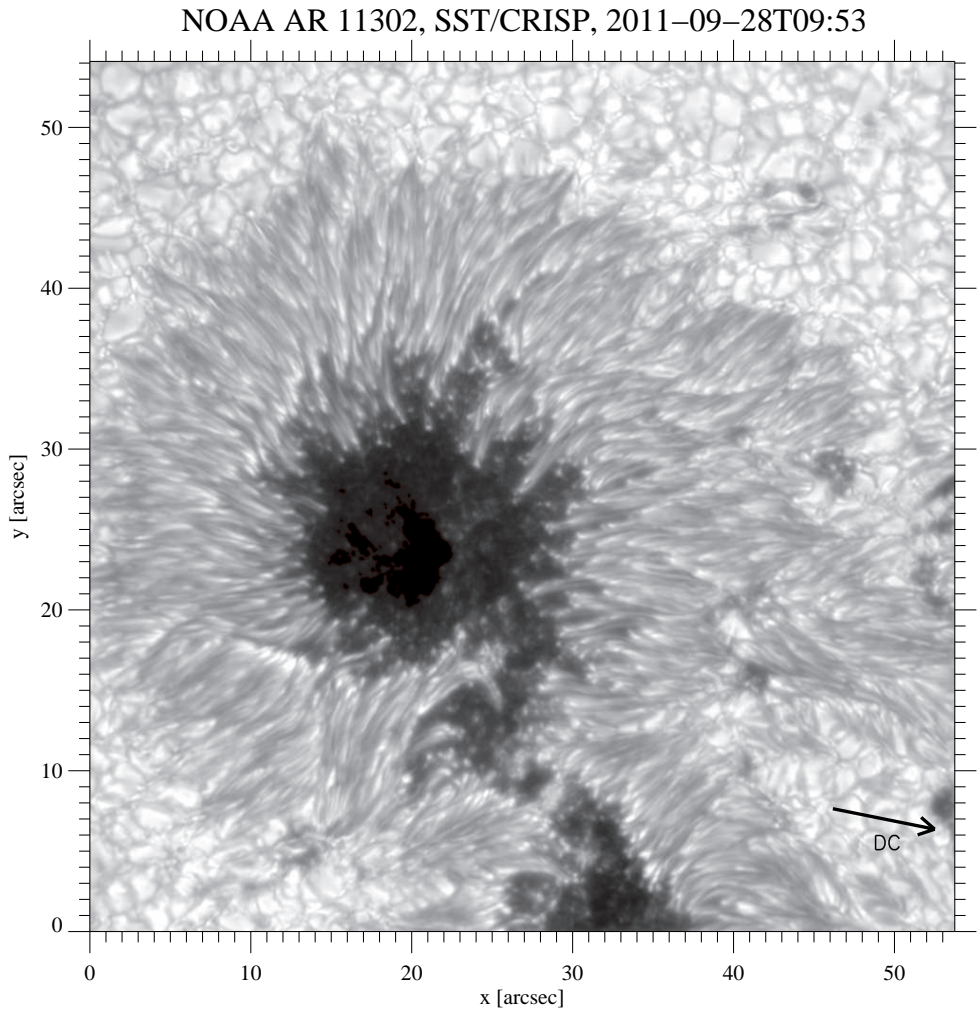}
        \caption{Continuum intensity image of the main sunspot of AR
        11302 as observed through the narrow-band channel of CRISP at
        an heliocentric angle of 6.8$^{\circ}$ on 28 September 2011,
        09:53 UT. The arrow points to the disk center.}
        \label{fig:continuo_dcarrow}
       \end{center}
\end{figure*}

However, the Evershed flow may not be the only manifestation of
convection in sunspot penumbrae.  \citet{2006A&A...460..605S} and
\citet{2006A&A...447..343S} proposed that penumbral filaments are
field-free gaps where regular convection takes place. Such a
  gappy penumbral model was later revised to include magnetic
  fields \citep{2008ApJ...677L.149S}. The velocity
field in this model consists of upflows at the center of the filaments
and downflows near their edges.  Three-dimensional MHD simulations of
penumbral fine structure \citep{2007ApJ...669.1390H,
  2009ApJ...691..640R, 2009Sci...325..171R, 2011ApJ...729....5R,
  2012ApJ...750...62R} support this scenario \citep[see the
  review by][]{2011LRSP....8....3R}. According to the simulations, hot
  gas rising from below the surface is deflected by the inclined
  magnetic field of the penumbra. This produces a fast flow toward the
  sunspot border---the Evershed flow.  Part of the rising gas turns
  over laterally and dips down below the solar surface, much in the
  same way as in quiet Sun granules and intergranular lanes.  Thus,
  the picture favored by the simulations is one of overturning
  convection. The convective cells are elongated in the preferred
  direction imposed by the magnetic field (the radial direction),
  forming penumbral filaments with a fast Evershed outflow along their
  axes and weaker downflows sideways.

An unambiguous confirmation of this picture---and hence of the
convective mechanism operating in the penumbra---requires the
detection of lateral downflows at the edges of the filaments.
However, the search has proven difficult.  Many authors studied the
penumbral velocity field without obtaining indications of downward
flows near the filament edges \citep[e.g.,][]{2003A&A...410..695M,
  2004A&A...427..319B, 2005A&A...436..333B, 2010ApJ...725...11B,
  2009A&A...508.1453F, 2010mcia.conf..186I, 2011PhDT.......137F}.
Most of those observations were made at intermediate angular
resolution, so they might have missed narrow structures.  Using
higher resolution data acquired with imaging spectrometers,
\citet{2011Sci...333..316S} and \citet{2011ApJ...734L..18J} reported
the detection of lateral downflows in penumbral filaments. In both
cases, a single snapshot of a sunspot at a relatively large distance
from the disk center was analyzed.  Moreover, the images were strongly
deconvolved to bring into view the weakest flow features hidden by
stray light arising from turbulence in the Earth's atmosphere.
The need for deconvolution together with the uncertainties in
  the parameters used to deconvolve the data cast doubts about the
  reliability of the inferred downflows. After the initial
discovery, other studies reported the detection of lateral downflows
\citep{2012A&A...540A..19S, 2013A&A...553A..63S, 2013A&A...549L...4R}.
By now these structures have been observed not only in imaging
data, but also in spectropolarimetric measurements from space.  In all
cases, it has been necessary to account for stray-light contamination
or the telescope point-spread function (PSF) to identify them with certainty. For example,
\cite{2013A&A...549L...4R} inverted Hinode spectropolarimetric data
deconvolved with the telescope PSF, while \citet{2013A&A...557A..25T}
used a spatially coupled inversion in which the synthetic Stokes
profiles were convolved with the Hinode PSF before being compared with
the observations. These detections are based on single snapshots too.

Here we study the small-scale velocity field of a sunspot penumbra
located close to the disk center, using a very stable time sequence of
high resolution spectropolarimetric measurements taken at the Swedish
1 m Solar Telescope \citep[SST;][]{2003SPIE.4853..341S}. The observations show
weak but ubiquitous lateral downflows at the edges of penumbral
filaments.  They are visible without performing any deconvolution of
the data. For the first time, the temporal evolution of these features
is observed and characterized. This allows us to understand why many
previous analyses failed to detect them.  The spatial and temporal
evolution of the small-scale velocity field reveals a convective
pattern similar to that prevailing in the quiet Sun, the main
difference being the existence of a strong radial Evershed
flow.

\section{Observations and Data Reduction}

\label{sec:observationsanddatareduction}

\begin{figure}[t]
\begin{center}
\includegraphics[trim=1cm 0cm 0cm 0cm, clip=true,width=.48\textwidth]{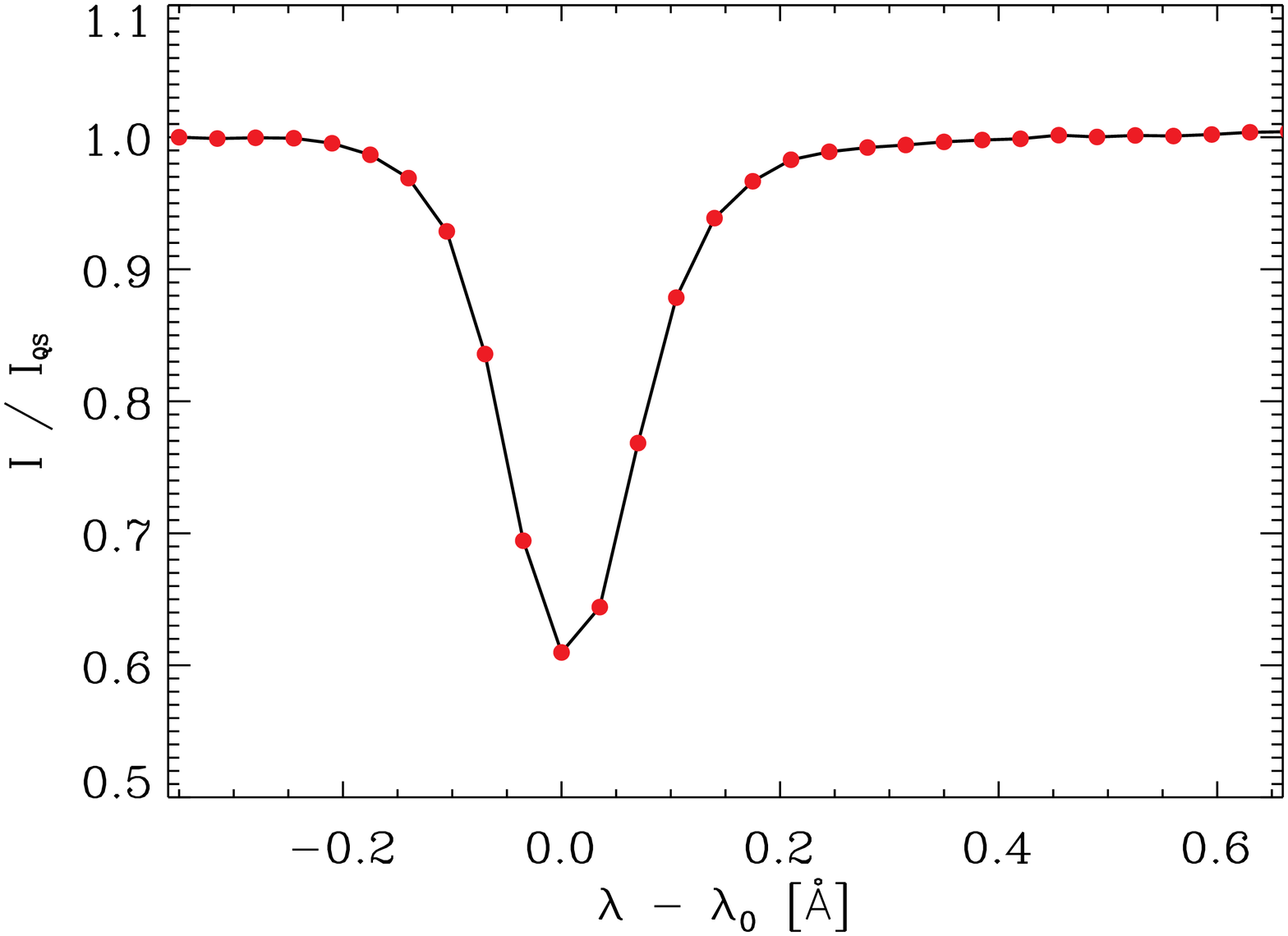}
\caption{Mean quiet Sun intensity profile of the \ion{Fe}{1} 617.3~nm
  line in the scan taken at 09:34:00~UT. The spectral sampling is shown with
  red circles. The $x$-axis represents the calibrated wavelength scale
  using the umbra as a reference.}
\label{fig:muestreo_linea}
\end{center}
\end{figure}

The data used here were obtained on 28 September 2011 between 09:20:40
and 10:03:52~UT. We followed the main sunspot of active region AR11302
under excellent seeing conditions (Figure~\ref{fig:continuo_dcarrow}).
At an heliocentric distance of only 6.8$^\circ$, the spot was located
very close to the disk center. Therefore, projection effects are
minimized and the line-of-sight (LOS) velocity mostly represents vertical motions.

The observations consist of time sequences of full Stokes measurements
in the photospheric \ion{Fe}{1} 617.3~nm line. They were acquired with
the CRisp Imaging Spectro-Polarimeter
\citep[CRISP;][]{2006A&A...447.1111S} at the SST on La Palma
(Spain).  CRISP is a dual Fabry-P\'erot interferometer capable of
delivering nearly diffraction-limited observations with the help of
the SST adaptive optics system \citep{2003SPIE.4853..370S} and the
Multi-Object, Multi-Frame Blind Deconvolution image restoration
technique \citep[MOMFBD;][]{2005SoPh..228..191V}.

Since we are interested in detecting lateral downflows that may be
narrow and weak, our aim was to secure very high spatial resolution
observations and high sensitivity to velocities. The diffraction limit
of the SST at 617~nm is $\lambda/D = 0\farcs13$, or 90~km on the Sun. On the other
hand, we chose the \ion{Fe}{1} 617.3~nm line because it is a narrow
photospheric transition suitable for Doppler shift measurements, has a
clean continuum, and do not show conspicuous line blends or telluric
lines nearby.

The \ion{Fe}{1} 617.3~nm line was sampled at 30 wavelength positions,
covering the range from $-35.0$ to $+66.5$~pm in steps of 3.5~pm
(Figure ~\ref{fig:muestreo_linea}). One spectral scan was completed in
32~s. We performed 9 acquisitions per modulation state, obtaining 36
images per wavelength position. The exposure time of the individual
images was 17~ms as set by an optical chopper. The full field of view
(FOV) of the observations covers an area of $55\arcsec \times 55
\arcsec$ with a plate scale of 0\farcs059 per pixel.

A very precise data reduction is crucial to this study, to avoid
identifying spurious velocities induced by instrumental errors as
lateral downflows. With that in mind we used the new CRISPRED pipeline
\citep{2015A&A...573A..40D}. Most of the complications involved in the
data reduction arise from corrugations in the surface of the CRISP
etalons, that could not be made infinitely flat. CRISP is mounted in
telecentric configuration very close to one of the focal planes,
comparatively optimizing image quality over a collimated configuration
\citep[see comments by][]{2006A&A...447.1111S}. In telecentric
configuration, however, the corrugations (cavity errors) produce
field-dependent random wavelength shifts of the CRISP
transmission profile, which effectively displace the observed line
profile on a pixel-by-pixel basis.

Current implementations of image reconstruction techniques assume that
there is an object that does not change among different realizations
of the seeing. Therefore, any intensity fluctuation that is detected
is assumed to be produced by atmospheric aberrations. In the presence
of a spectral line profile, cavity errors introduce a pattern of
intensity fluctuations that are not caused by seeing motions. Without
any compensation for these fluctuations, the resulting reconstructed
images can contain small-scale artifacts that correlate with the
cavity-error map. This problem usually appears when seeing 
conditions are not great and/or the line profile is very steep.

As an approximate solution, \citet{2011A&A...534A..45S} proposed to
correct these fluctuations assuming a quiet-Sun spectral profile. This
assumption is somewhat valid on granulation observations, but it is
not precise enough in sunspots where the strength and width of the
profile substantially deviate from quiet-Sun values. Also, the
\ion{Fe}{1}~617.3~nm line used here is particularly steep and narrow,
compared with, e.g., the \ion{Fe}{1}~630~nm lines that have been
commonly employed in other photospheric studies.

A much more accurate solution to this problem has been given by
\citet{2015A&A...573A..40D}. They improved the flat-fielding of the
data prior to image reconstruction in two ways. First, the quiet-Sun
profile that is present in the flats is removed
\citep{2011A&A...534A..45S}. To compensate for the intensity
fluctuations, all the acquisitions corresponding to the same polarization state and
same wavelength within a line scan are summed, resulting in a
(somewhat) spatially blurred image of the object with dimensionality
$(x,y,\lambda)$.  Those spectra are used to estimate the imprint of
the cavity error in each pixel, by shifting the line profile to the
real observed grid.  A linear correction is then introduced in each
pixel for each wavelength, polarization state, and time step.

Furthermore, \citet{2015A&A...573A..40D} argue that, since the summed
images may still contain distortions from the seeing, it is better to
compensate only for the high frequencies present in the cavity map,
which are associated with the artifacts produced by cavity errors.

This strategy minimizes the intensity variations due to cavity errors,
and also allows to estimate the shift of the wavelength scale on a
pixel-by-pixel basis, taking into account how the image is changed by
the MOMFBD restoration.

The polarimetric calibration was performed for each pixel as described
in \citet{2008A&A...489..429V}. Telescope-induced polarization was
corrected using a theoretical model of the SST
\citep{2010arXiv1010.4142S} with updated parameters for the 617~nm
spectral range. Residual seeing within each line scan and wavelength
was removed following \citet{2012A&A...548A.114H}, so all polarization
states were aligned with subpixel accuracy before demodulating the
data. Finally, the resulting time series of Stokes images were
de-rotated, co-aligned and de-stretched (to compensate for
rubber-sheet motions) as described in \citet{1994ApJ...430..413S}.

\section{Data Analysis}
\label{sec:dataprocessing}

\subsection{Determination of LOS Velocities}
\label{calculatinglosvelocities}

We construct LOS velocity maps using line bisectors. The bisector is
the curve dividing a spectral line into two halves. It is computed by
finding the midpoints of horizontal cuts of the line at different
intensity levels, from the core (0\%) to the continuum (100\%).  Bisectors
allow the depth dependence of the velocity field to be traced in each
pixel, as intensity levels close to the continuum sample deeper layers
than those near the line core.  By mapping a given bisector level
across the FOV, we are measuring velocities from a corrugated
geometrical surface on the solar atmosphere.

Bisectors are calculated from the observed intensity profiles using
linear interpolation, after spectral gradients due to the CRISP
prefilter have been removed. In this paper we focus on the bisectors
at the 50, 60 and 70\% levels, to progressively sample deeper
layers of the photosphere in each pixel. The Evershed flow is stronger
close to the continuum forming layer \citep[][and
others]{1995A&A...298..260R, 2001ApJ...547.1148W,
  2003ASPC..307..301B}, so in principle bisectors at high intensity
levels are advantageous to detect the weak signals expected from
lateral downflows.

We have removed the 5-minute photospheric oscillations
\citep{1962ApJ...135..474L} from the velocity maps using a Fourier
filter with cut off speeds of 5 km~s$^{-1}$ \citep{1989ApJ...336..475T, 
1992A&A...256..652S}.

\subsection{Velocity Calibration}
\label{subsec:calibration}

The filtered Dopplergrams were calibrated using the umbra as a velocity
reference \citep{1977ApJ...213..900B}. To avoid possible molecular
blends, we selected all umbral pixels ($I_{\rm c} < 0.45 I_{\rm QS}$) with
Stokes V amplitude asymmetries\footnote{The amplitude asymmetry is
  defined as $\delta a = (a_{\rm b}- a_{\rm r})/(a_{\rm b}+ a_{\rm
    r})$, where $a_{b}$ and $a_{r}$ represent the amplitudes of the
  Stokes V blue and red lobes, respectively. We calculated them
  fitting a parabola to three points around the maximum of each lobe.}
smaller than 4\%. This criterion is met by approximately 60\% of the
umbral points.

For each Dopplergram in the sequence, the zero point of the velocity
scale was taken to be the average Stokes V zero-crossing point of the
umbral pixels selected in that frame. The typical standard deviation
of the zero-crossing points used to compute the velocity reference is
110 m~s$^{-1}$.  This can be considered the uncertainty of our
velocity calibration. The standard deviation of the average Stokes~V
zero-crossing points in the different line scans is 3~m~s$^{-1}$.

\section{Results}
\label{sec:analysis}
In this Section we investigate the penumbral velocity field on all
spatial scales accessible to our observations. We first
  describe the Evershed flow as seen in the Dopplergrams. Then we
  summarize the properties of narrow, elongated downflows that occur at the edges
  of the penumbral filaments. In the third part of this section we show
the evolution of a few representative examples of such downflows.
Finally, we present a statistical study of their physical properties
in the center-side penumbra.

Figure~\ref{fig:dopplergrama_grande_paper} shows one of the
Dopplergrams of the temporal sequence and an enlargement of a
center-side penumbral region. The Dopplergram corresponds to the 70\%
bisector level and samples deep layers of the solar atmosphere. The
contours outline the inner and outer penumbral boundaries. The black
arrow points to the disk center. A movie of the Dopplergram sequence
can be found in the electronic Journal.

\subsection{Evershed Flow}
\label{subsec:description}

\begin{figure*}[t]
	\begin{center} 
	\includegraphics[trim = 2.5cm 0cm 6cm 0cm, clip =
	true, width = 1.\textwidth]{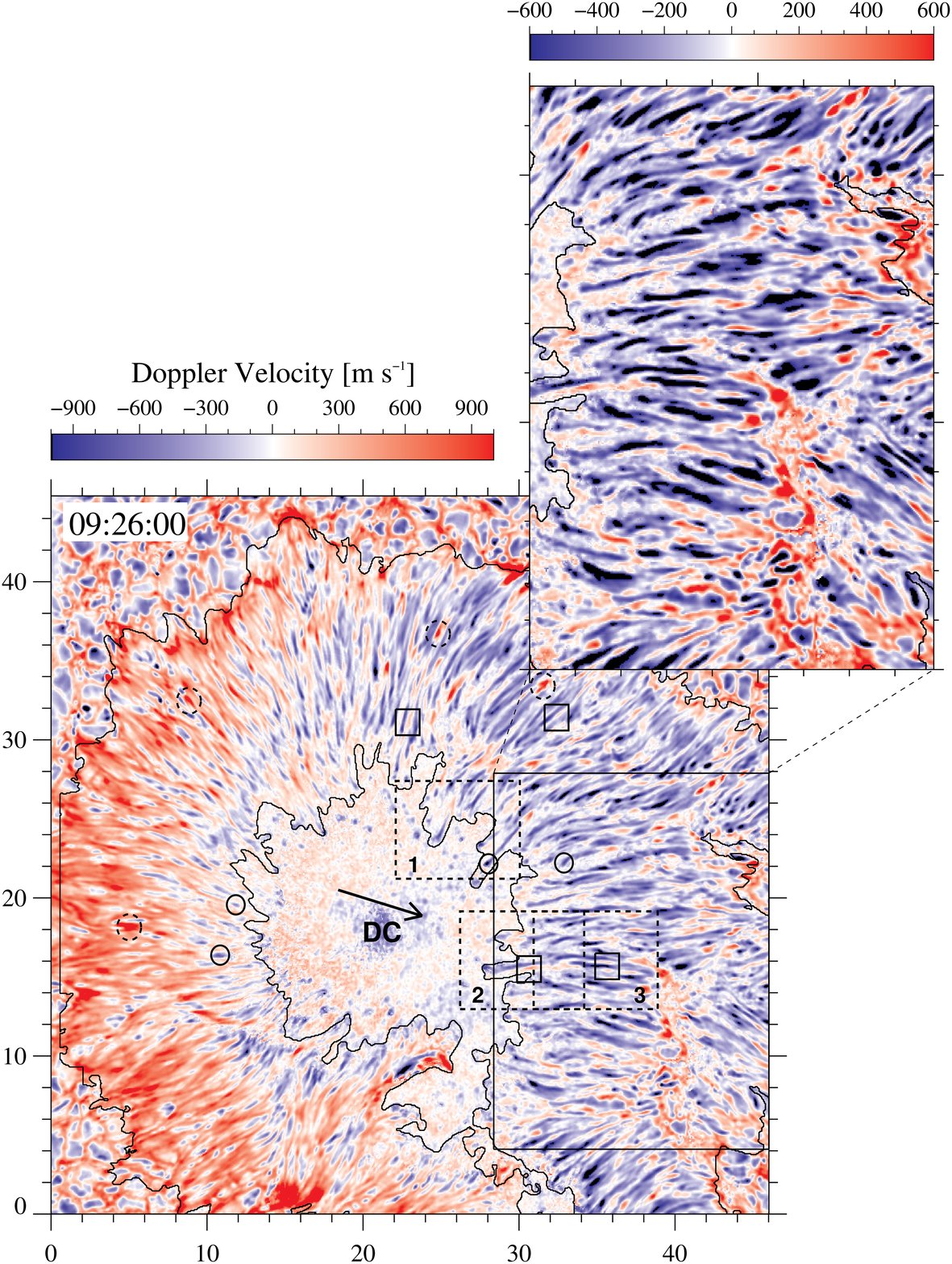}
	\caption{Velocity field in and around AR 11302 as
          derived from the line bisectors at the 70\% intensity level.
          Both axes are in arcsec.  The arrow points to the
          solar disk center and contours outline the inner and outer
          boundaries of the penumbra. Solid circles show
            examples of penumbral grains, dashed circles mark strong
            returning Evershed flows, and small squares enclose
            lateral downflows.  The dashed rectangles labeled 1--3
            identify the areas used in
            Figures~\ref{fig:example33}, \ref{fig:example40} and
            \ref{fig:example7}. A region of the center-side penumbra
            is enlarged and displayed with a different scale in the
            right panel to better illustrate the existence of lateral
            downflows at the edges of penumbral filaments.}
	\label{fig:dopplergrama_grande_paper}
	\end{center}
\end{figure*}

Despite the small heliocentric angle of the sunspot, the typical
Evershed pattern is clearly seen in
Figure~\ref{fig:dopplergrama_grande_paper}. The limb-side and
center-side penumbrae are redshifted and blueshifted, respectively,
indicating motions away from and to the observer. These motions become
stronger toward the outer sunspot boundary. The observed pattern of
Doppler shifts demonstrates that the penumbral flow is, to first
order, a radially directed outflow whose inclination to the vertical
steadily increases toward the outer penumbra. This picture does not
differ much from previous results \citep[e.g.,
][]{1997ApJ...477..485S, 2000A&A...358.1122S, 2003A&A...410..695M,
2004A&A...422.1093B, 2004A&A...427..319B, 2006ApJ...646..593R}.
However, the very high spatial resolution of our observations allows
us to identify individual flow channels with unprecedented accuracy.
Elongated, extremely thin flow structures are observed to stretch
radially outward for a few arcseconds in the velocity map. They
exhibit opposite velocities on the two sides of the spot.  These
structures often---but not always---coincide with bright penumbral
filaments.

The flow channel heads display conspicuous, localized patches of
enhanced blueshifts regardless of their position in the spot, although
they are easier to detect in the inner penumbra because that region is
less cluttered. The solid circles in
Figure~\ref{fig:dopplergrama_grande_paper} mark prominent examples.
The strong blueshifts represent upward motions and can be considered
to be the sources of the hot Evershed flow.  They always occur at the
position of bright penumbral grains \citep{2006ApJ...646..593R,
  2007PASJ...59S.593I}. Our observations show them with typical sizes
of 0\farcs2, in agreement with previous studies.

The tails of the flow channels, especially those near the outer
penumbral edge, often display strong redshifts.
These patches represent the locations where the Evershed flow returns
to the solar surface \citep[e.g.,][]{1997Natur.389...47W,
  2001ApJ...547.1148W}, now observed with unprecedented clarity as
separate structures. They can be identified most easily in the
center-side penumbra, thanks to their large contrasts over the blue
background. Examples are marked with dashed circles in
Figure~\ref{fig:dopplergrama_grande_paper}. The downflows occurring in
these patches are supersonic and nearly vertical
\citep{2010mcia.conf..193B, 2013A&A...557A..24V}, although the
bisector analysis yields modest LOS velocities of order 2~km~s$^{-1}$.

\subsection{Lateral Downflows}
\label{sec:interpretation}
In addition to these flow structures, our Dopplergrams reveal the
existence of a new flow component producing small patches of weak
redshifts in between the flow filaments. They are observed everywhere
across the spot, except in the limb-side penumbra where the prevailing
redshifts hide them efficiently. By contrast, the patches are very
conspicuous in the center-side penumbra and the region perpendicular
to the symmetry line (the line connecting the sunspot to the disk
center).  Examples of such lateral redshifts are indicated in
Figure~\ref{fig:dopplergrama_grande_paper} with small squares.  Their
general properties can be summarized as follows:
\begin{itemize}
\item They are associated with flow filaments. Sometimes the lateral
  redshifts can be observed on both sides of the filament, but most
  often they appear only on one side.
\item Their sizes vary from small roundish patches $\sim$0\farcs15 in
  diameter to elongated, narrow structures flanking the flow filaments
  for more than 1\arcsec.
\item In the center-side penumbra, they show bisector velocities
  ranging from $\sim$100 to $\sim$500 m~s$^{-1}$ at the 70\% intensity
  level. These flows are therefore much weaker than those observed in
  the adjacent filaments.
\item Their velocities are stronger in deeper photospheric layers.  In
  the center-side penumbra, they exhibit mean maximum velocities of
  165, 195 and 210~m~s$^{-1}$ at the 50, 60 and 70\% intensity
  levels, respectively.
\end{itemize}

What do these flows represent? The Doppler velocities displayed in
Figure 3 are projections of the actual velocity vector to the LOS.
Along the symmetry line, only radial and vertical motions can
contribute to the LOS velocity.  The center-side penumbra near the
symmetry line is dominated by the blueshifts produced by the radial
Evershed outflow. Any redshift observed there must be due to vertical
downdrafts or to inward horizontal motions. The latter can be
ruled out because the structures move outward rather than inward
(see below). Thus, the lateral redshifts must correspond to 
downward vertical flows.

In the region perpendicular to the symmetry line, only
  horizontal flows along the azimuthal direction (i.e., perpendicular
  to the radial direction of the filaments) and vertical flows can
produce non-zero LOS velocities.  The global reduction of the LOS
velocity observed there is caused by the vanishing contribution of the
radial Evershed flow.  In that part of the spot, however, we still
detect small redshifted patches in between the penumbral filaments.
Since they often occur in pairs on either side of the same flow
channel, they cannot be due to horizontal motions away from the
filament, as those motions would produce velocity patches of opposite
sign. In addition, due to the small heliocentric angle of the spot,
one would need very large azimuthal flows to explain the observed LOS
velocities.  Therefore, also in this region we conclude that the
redshifted patches represent vertical motions away from the observer,
i.e., downflows.

\begin{figure}[t]
\begin{center}
\includegraphics[trim = 3.1cm 1cm 1cm 0.5cm,clip=true,width = 0.52\textwidth]{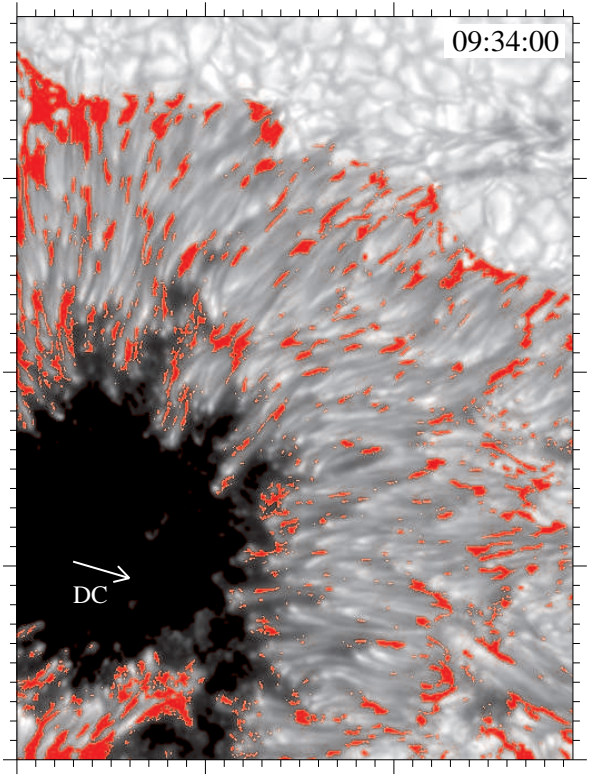}
\caption{Spatial distribution of downflows in the penumbra of AR
    11302. Red pixels represent locations with bisector velocities
  larger than 100~m~s$^{-1}$ at the 70\% intensity level, overplotted
  on the continuum intensity filtergram.}
\label{fig:down_continuo_new_25}
\end{center}
\end{figure}

Figure~\ref{fig:down_continuo_new_25} shows the location of the
downflows relative to the intensity structures.  On top of the
continuum intensity filtergram, red pixels indicate bisector
velocities larger than 100 m~s$^{-1}$ at the 70\% intensity level. As
can be seen, the lateral downflows tend to appear next to penumbral filaments, but sometimes they occur at their edges
or even right on them.  As a consequence, the lateral downflows are
not always associated with dark structures.

\begin{figure}[t]
	\begin{center}
    	\includegraphics[trim = 1cm 1cm 0cm 1.5cm, clip = true, 
         width = 0.48\textwidth]{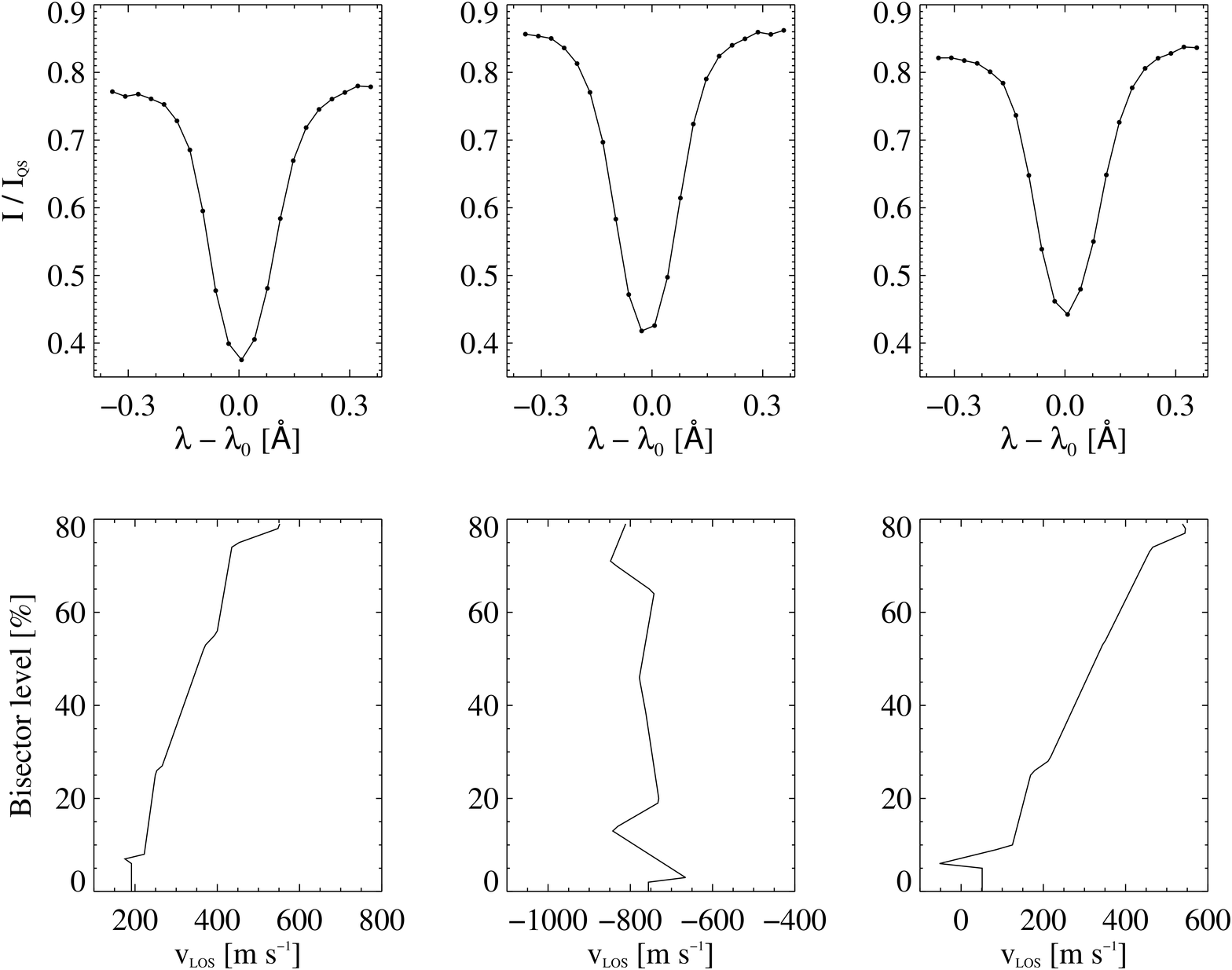}
  	\caption{Top row: intensity profiles emerging from a
          blueshifted flow channel (central panel) and its lateral
          downflowing regions (right and left panels). The positions
          of the selected pixels are marked in
          Figure~\ref{fig:example7} with plus symbols. Bottom row:
          line bisectors calculated between the 0 and 80\% intensity
          levels. }
  		\label{fig:perfiles_33_15}
  		\end{center}
\end{figure}

Figure~\ref{fig:perfiles_33_15} displays the intensity profiles
emerging from a blueshifted penumbral filament (central panel) and its
two lateral downflows (right and left panels).  The exact location of
the profiles is marked 
with plus symbols in the close-up of Figure~~\ref{fig:example7}. Also
depicted are the corresponding line bisectors for intensity levels
from 0 to 80\%.  As can be seen, the intensity profiles emerging from
the lateral downflows have slightly darker continua than the flow
channel.  Other properties such as line widths are similar.  The
lateral downflows produce bisectors with larger redshifts at higher
intensity levels, indicating the existence of velocities that increase
with depth in these pixels. The bisector at the position of the flow
channel reveals upflows of order 800~m~s$^{-1}$ that do not change
much throughout the atmosphere.  These spectral properties reveal the
different nature of the central upflows along the filament and the
lateral downflows.

\subsection{Temporal Evolution of Lateral Downflows}
\label{subsec:behavior_individual}

\begin{figure*}[t]
	\begin{center} 
	\includegraphics[trim = 2.25cm 2.25cm 2.5cm 0cm, clip
	= true, width
	=1.\textwidth]{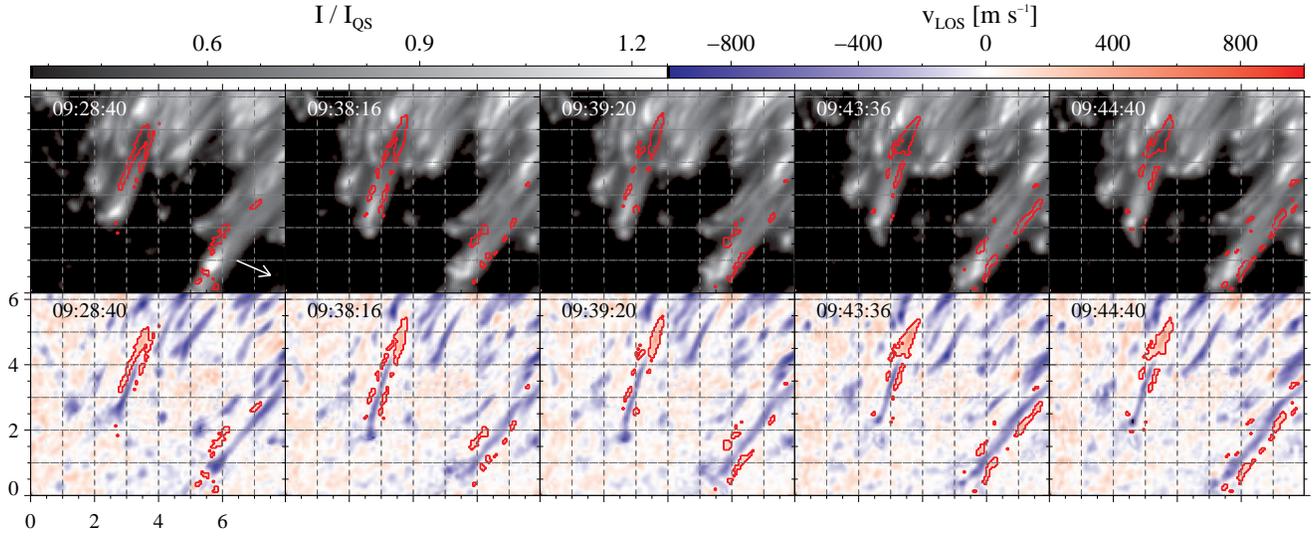}
	\caption{Temporal evolution of lateral downflows in penumbral
	filaments located perpendicularly to the symmetry line (case
	1). The intensity panels on top have been corrected for a
	stray light contamination of 40\% (see
	Sect.~\ref{sub:straylight}). The velocity maps in the second
	row display the bisector velocities at the 70\% intensity
	level without correction. Red contours outline redshifts greater than
	100~m~s$^{-1}$.  The frames are not contiguous in time, as
	indicated in the upper right corner of each panel. The white
	arrow plotted in the first intensity panel points to the disk
	center. Axes are labeled in arcsec.}  \label{fig:example33}
\end{center}
\end{figure*}

\begin{figure*}[t]
	\begin{center}
	\includegraphics[trim = 2.25cm 2.25cm 2.5cm 0cm, clip
	= true, width =
	1.\textwidth]{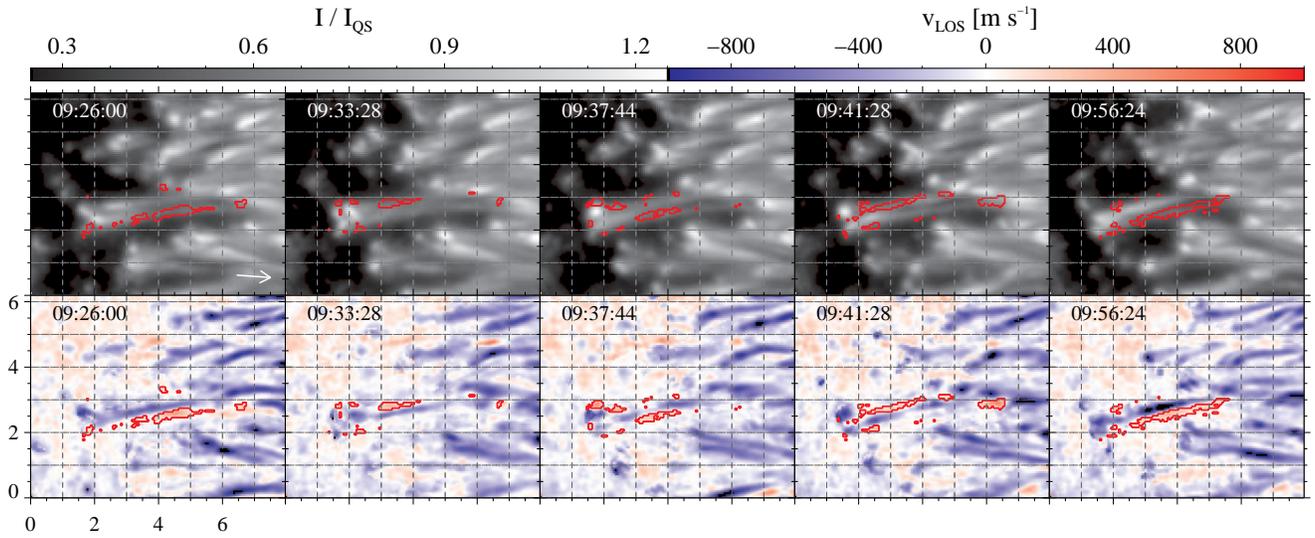}
	\caption{Example of lateral downflows located at the edges of
	a conglomerate of filaments protruding into the umbra near the
	symmetry line (case 2). The layout is the same as in
	Figure~\ref{fig:example33}.}  \label{fig:example40}
\end{center}
\end{figure*}

\begin{figure*}[t]
	\begin{center}
	\includegraphics[trim = 3.50cm 2.25cm 4.5cm 0cm, clip
	= true, width =
	1.\textwidth]{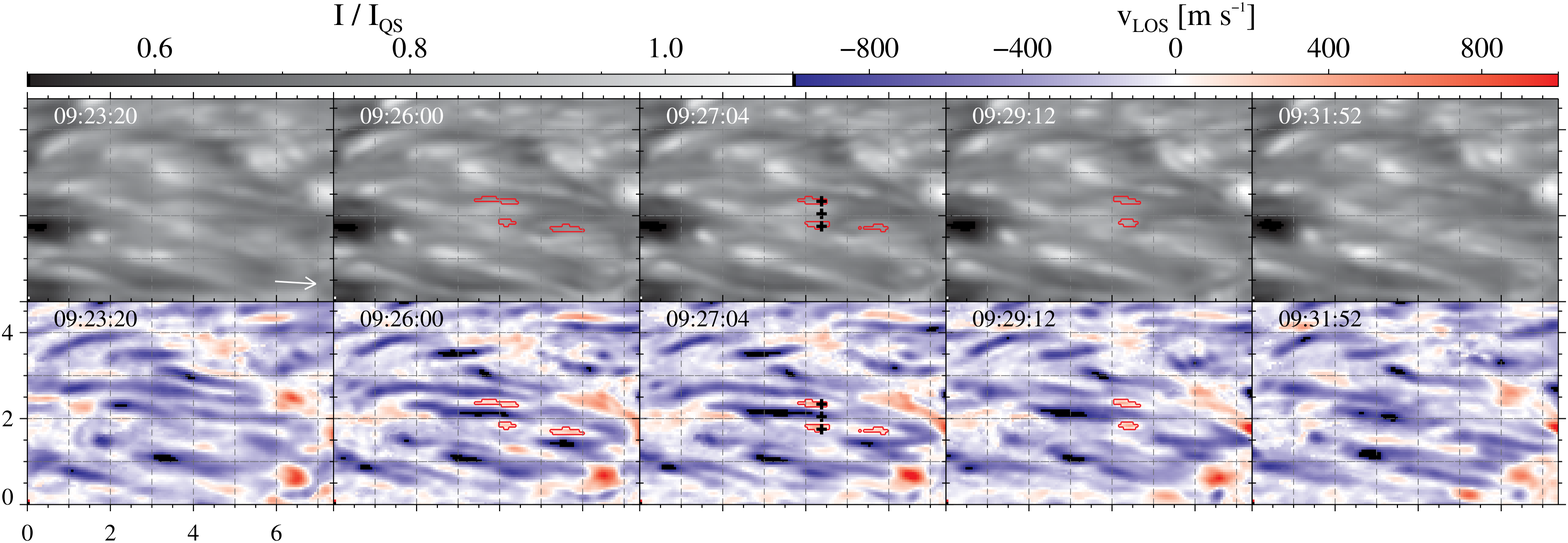}
	\caption{Lateral downflows at the edges of a filament located
          in the middle penumbra near the symmetry line (case 3). The
          layout is the same as in Figure~\ref{fig:example33}. The
          plus symbols at 09:27:04 UT mark the pixels represented in
          Figure~\ref{fig:perfiles_33_15}.}  \label{fig:example7}
          \end{center}
\end{figure*}

We have selected three examples to illustrate the evolution of the
lateral downflows (Figures~\ref{fig:example33}, ~\ref{fig:example40},
and ~\ref{fig:example7}). They are located at different positions
across the center-side penumbra (dashed rectangles in
Figure~\ref{fig:dopplergrama_grande_paper}). Movies covering their
entire evolution are available in the electronic Journal.

\subsubsection{Case 1}

Figure~\ref{fig:example33} shows a region perpendicular to the
symmetry line in the inner penumbra. Two bright filaments protruding
into the umbra can be seen in the intensity maps (top row). They
appear in the Dopplergrams (bottom row) as two blueshifted flow
channels headed by roundish patches that also harbor blueshifts. Since
the filaments are nearly perpendicular to the symmetry line, the LOS velocity is the result of azimuthal and vertical
motions.  Following the arguments of Section~\ref{subsec:description},
these blueshifts represent the upward component of the Evershed flow
in the penumbra, with velocities of up to 800~m~s$^{-1}$.

Small redshifted patches can be seen next to the two penumbral
filaments. We have identified them with red contours.  They never
cover the entire length of the filaments, but appear as patchy, more
or less elongated structures close to their edges. In the intensity
images, they are located next to bright structures or even superposed
on them. Sometimes we see the redshifts only on one side of the
filament, sometimes on both sides, evolving independently of each
other.  Given their location, the redshifts are unlikely to be
produced by azimuthal motions and must rather be considered as
downward motions.

The lateral downflows depicted in Figure~\ref{fig:example33} are
intermittent. Individual patches usually have lifetimes of a few
minutes, although the bigger ones are recognizable for up to
10~minutes.  Their velocities remain more or less
stable with time, but they can split into smaller fragments and/or
merge with neighboring patches from one frame to the next. These
processes may simply be the result of an inhomogeneous distribution of
downward velocities along the filament length, with regions of
enhanced flows which are easy to detect and areas of weaker flows that
can go largely unnoticed.  If the velocities change with time one may
get the impression that the patches split and merge, while in reality
there exists only one long patch covering the full filament length. An
inhomogeneous velocity distribution can also result in apparent drifts
of the strongest downflowing patches, as is often observed.

\subsubsection{Case 2}

Our second example is shown in Figure~\ref{fig:example40}. Now, the
filaments are located nearly parallel to the symmetry line and the
LOS velocity is mostly due to radial and vertical motions.
We see a conglomerate of filaments protruding into the umbra.  Small
patches of redshifts appear and evolve individually around the most
prominent one. Again, these redshifts in the center-side penumbra must
be caused by downward motions.

The lateral downflows of Figure~\ref{fig:example40} are also
intermittent, because they appear and disappear with time. They
interact between them, merging and splitting in smaller parts as
described before.  However, the downflowing patches do not evolve in a
completely independent way here. In the first frame we observe the
blueshifted channel above a bead-like string of lateral downflows of
different sizes.  The blueshifted region moves downward (in the plane
of the paper), interchanging position with the lateral downflows in
the second frame.  There is a similar transition between the fourth
and fifth frames, as the blueshifted filament moves upward and
interchanges position with the redshifted patches that are mostly seen
again on the other side. Therefore, in this example the lateral downflows appear to
follow---or react to---the motion of the flow filament.

\subsubsection{Case 3}

Our last example corresponds to a mid-penumbral region where the
filaments are well aligned with the symmetry line
(Figure~\ref{fig:example7}).  The Dopplergrams show elongated flow
channels harboring blueshifts.  These structures are difficult to
trace in the continuum images because of their low contrast and the
complexity of the region.  As in the previous example, the LOS
velocity represents the projection of the radial and vertical
components of the velocity vector, since the azimuthal component is
perpendicular to the LOS.  The selected region contains several
examples of weak lateral downflows, but we will focus on the ones
marked with contours.

Initially, no redshifts can be observed at the edges of the central
flow channel. In the second frame, three redshifted regions have
appeared.  The onset of the redshifts is perhaps triggered by a
significant velocity increase happening in the filament, from $-500$
to $-1000$~m~s$^{-1}$. Two of the patches are located on either 
side of the filament, approximately at the position of the strongest
blueshifts.  One of them is more elongated than the other. The third
patch is observed near the tail of the filament, but we believe it
belongs to a nearby flow channel. The lateral downflows move outward
between the second and third frames. By then, their sizes are
approximately the same and their velocities have increased from 250 to
400~m~s$^{-1}$.  They no longer flank the strongest blueshifts, which
have moved slightly inward. In the fourth frame, the redshifts are
further outward, close to the filament tail. Their areas are still
similar, but the velocities have decreased back to 250~m~s$^{-1}$. In
the last frame the redshifts have been replaced by blueshifts.  This
suggest that the development of blueshifts ultimately determines the
appearance and visibility of the downflowing patches.

\begin{figure*}[!t]
\begin{center}
\includegraphics[trim=6cm 0cm 2cm 0cm, width=1\textwidth]{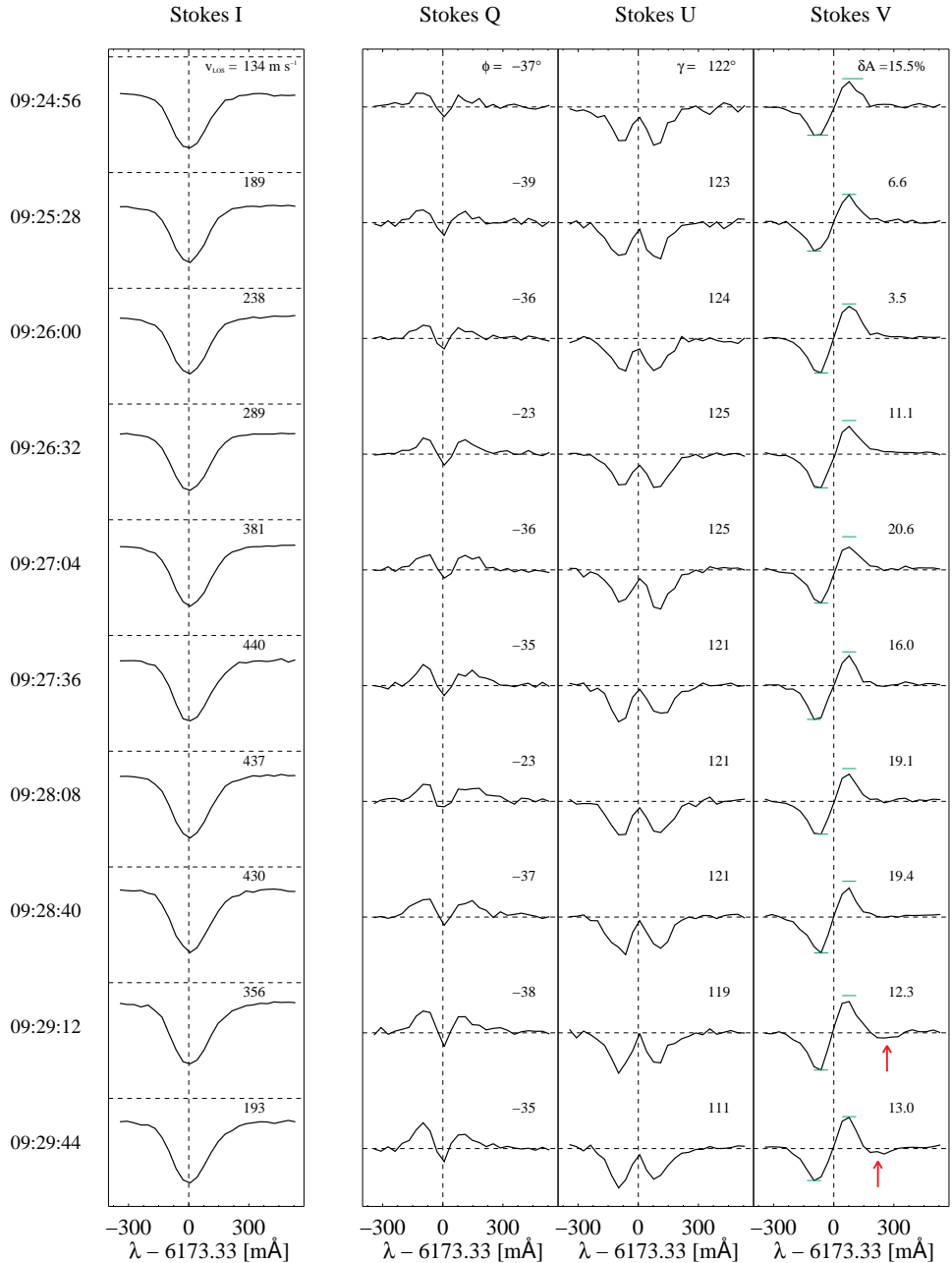}
\caption{Temporal evolution of the Stokes profiles emerging from the
  center of the lowermost lateral downflow in
  Figure~\ref{fig:example7}. Stokes I, Q, U and V are shown from left
  to right, and time increases from top to bottom, as indicated on the
  left hand side. The horizontal lines in the intensity panels
  represent the continuum of the average quiet Sun profile, making it
  possible to know if the structure is bright or dark by direct
  comparison with the observed continuum. The horizontal lines in the
  other columns indicate zero polarization signal. The small
  horizontal dashes in the last column represent the amplitudes of
  the Stokes V profiles. They allow the amplitude asymmetry to be quickly
  assessed by simple visual inspection. The numbers in the upper right
  corners of the panels indicate the Doppler velocity at the 70\%
  intensity level (first column), the magnetic field azimuth (second
  column), the magnetic field inclination (third column; 90$^\circ$
  represents fields perpendicular to the LOS and $180^\circ$ fields
  pointing away from the observer along the LOS), and
  the Stokes V area asymmetry in percent (fourth column). The magnetic
  parameters have been determined assuming complete Zeeman splitting
  as in \cite{2010ApJ...713.1282O}.}
  \label{fig:profiles}
  \end{center}
\end{figure*}

\subsection{Stokes Profiles}
In Figure~\ref{fig:profiles} we display the four Stokes profiles 
emerging from the lowermost lateral downflow marked with a contour in
Figure~\ref{fig:example7}, as a function of time. Both linear and
circular polarization are clearly seen. The most prominent spectral
signature of the profiles, however, is the noticeable asymmetry of
Stokes~I, which shows increasing redshifts toward the continuum (the
vertical lines indicate the rest position of the line).

The linear polarization spectra do not show features worth of mention,
except perhaps their amplitude asymmetries.  The circular polarization
profiles have positive area asymmetries, with the blue lobe being
more extended than the red lobe (the exact values are shown next to
the Stokes V profiles). In other patches we observe negative area
asymmetries. This indicates the existence of velocity gradients along
the LOS, possibly coupled with magnetic field gradients, which is
consistent with the information provided by the line bisectors.  Other
than that, the Stokes V profiles emerging from the redshifted patch
have the same sign as in the umbra. Thus, the magnetic field vector
does not appear to undergo a polarity reversal there.  However, toward
the end of the sequence, at 09:29:12 and 09:29:44~UT, the Stokes V
spectra exhibit a weak third lobe in the red line wing, close to the
continuum. This additional lobe (marked with arrows) seems to be
strongly redshifted and of opposite polarity compared with the main
Stokes V lobes.  Similar signals are observed in many pixels of the
redshifted patch at those times of the evolution. They occur when the
downflowing velocity is decreasing (as seen in the intensity profile),
but when the asymmetry of Stokes I is maximum, with a very extended
wing toward the red continuum.

Three-lobed profiles have previously been detected in the
  outer penumbra---both near the sunspot neutral line
  \citep{1992ApJ...398..359S, 2002A&A...381..668S} and far from it
  \citep{2001ApJ...547.1148W, 2002NCimC..25..543B}---, at the position
  of Evershed flows returning to the solar surface
  \citep[e.g.][]{2007PASJ...59S.593I, 2013A&A...550A..97F}, and at the
  edges of penumbral filaments \citep{2013A&A...549L...4R,
  2013A&A...553A..63S}. The mere detection of these profiles indicate
  the coexistence of fields of opposite polarity along the line of
  sight.  The opposite polarities do not need to be cospatial (i.e.,
  to lie side by side), as they can also be located at different
  heights in the atmosphere. Whether the third lobe is produced by
  dragging of the spot field lines by the downflows or by a stable
  field pointing to the solar surface remains to be investigated.

The direct detection of opposite polarities in the circular
polarization spectra may have occurred just by chance, as a result of
slightly stronger downflows coupled with more vertical fields leading
to a more favorable projection. These fields could be present all the
time in the redshifted patches, but without leaving clear signatures
in the spectra because of unfavorable conditions. We plan to carry out
a full Stokes inversion of the observations to determine whether
opposite magnetic polarities exist during the whole lifetime of the
patches, or only intermittenly (which would favor dragging and
deformation of existing field lines at the edges of penumbral
filaments).

\subsection{General Properties}

We have determined the properties of the lateral downflows through
repeated application of the YAFTA feature tracking code
\citep{2003ApJ...588..620W} to the disk-center penumbral region zoomed
in Figure~\ref{fig:dopplergrama_grande_paper}, which contains the symmetry line. Lateral downflows
are identified as structures with a minimum size of 5 pixels and LOS
velocities between 50~m~s$^{-1}$ and 600~m~s$^{-1}$ in the individual
velocity maps, using the clumping method. The upper limit for the
velocity excludes redshifts at the tail of penumbral filaments that
correspond to returning Evershed flows. After experimenting with
different thresholds, the YAFTA identifications were further refined
by retaining only those structures visible in at least four
consecutive frames (to avoid false detections due to noise) or coming
from the fragmentation of another structure.
The features detected through application of these criteria were
checked manually to ensure that they represent genuine lateral
downflows.  All in all, 754 structures (5328 patches) were 
identified and tracked in the Dopplergram time sequence.

\begin{figure}[t]

	\begin{center}
	\includegraphics[trim=0.8cm 2.85cm 5.3cm 5cm,clip =
	true, keepaspectratio = true, width =
	0.45\textwidth]{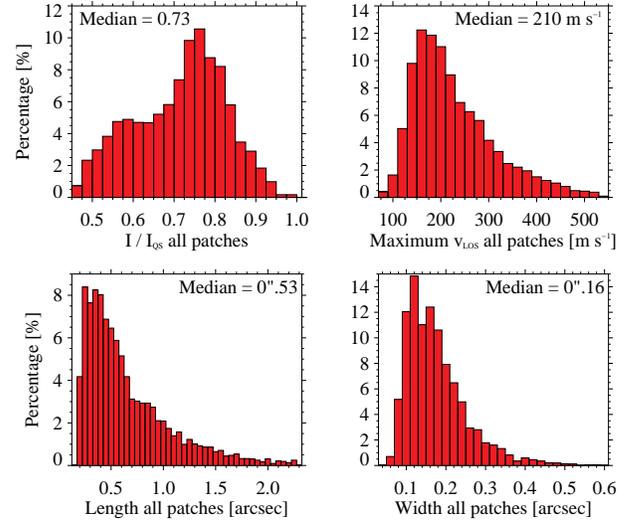}
	\caption{Histograms of continuum intensity,
	maximum LOS velocity, length, and width of the redshifted patches observed at the
	edges of penumbral filaments, considered individually. The
	median of the distribution is indicated in the upper right
	corner of the panels.}  \label{fig:histograma_downflows}
\end{center}
\end{figure}

\begin{figure}[t]

	\begin{center}
	\includegraphics[trim=0.8cm 3cm 5.3cm 5cm,clip =
	true, keepaspectratio = true, width =
	0.45\textwidth]{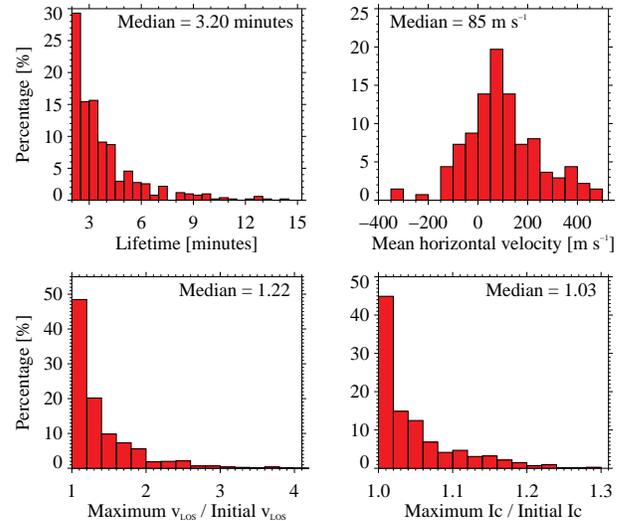}
	\caption{Histograms of lifetime, mean horizontal speed, ratio
          of maximum to initial continuum intensity, and ratio of
          maximum to initial LOS velocity for the lateral downflows,
          considered as coherent structures that evolve from frame to
          frame.  The median of the distribution is indicated in the
          upper right corner of the panels.}
        \label{fig:histograma_downflows2}
        \end{center}
\end{figure}

Figure~\ref{fig:histograma_downflows} displays histograms of the
continuum intensity, maximum LOS velocity, length, and width of the
individual patches detected by YAFTA. The lateral downflows show a
broad range of intensities, indicating that they appear both
in dark regions outside the filaments and in the bright filament
edges. The median value is 0.73 of the quiet Sun continuum
intensity. Most patches have a maximum LOS velocity in the range
150--250~m~s$^{-1}$ with a median of 210~m~s$^{-1}$. We have fitted an
ellipse to each patch to obtain its dimensions.  The major axis of the
ellipse corresponds to the length of the patch and has a median value
of 0\farcs53.  The width is given by the minor axis, which varies
between $\sim$0\farcs1 and $\sim$0\farcs3 with a median of
0\farcs16. Therefore, the redshifted patches tend to be elongated and
very narrow.

Figure~\ref{fig:histograma_downflows2} summarizes the properties of
the lateral downflows, considered as coherent structures that can be
tracked in subsequent frames of the Dopplergram sequence.  We show
histograms of lifetimes, horizontal motions, and the ratios of maximum
to initial intensities and LOS velocities. For the calculation of the
lifetime we have considered only the structures that appear and
disappear in situ (i.e., that do not undergo interactions) and those
that appear in situ and fragment, but only until their first
fragmentation. The fragments are excluded in order not to bias the
statistics, since they usually have shorter lifetimes. For this reason
it is appropriate to think of the lifetimes as detection times.  Most
of the lateral downflows can be observed for 4--10 frames. The
distribution decreases exponentially, the median being 3.2 minutes (6
frames). Thus, they are short-lived features.  This may be 
intrinsic to the physical mechanism driving the downflows or a 
consequence of our inability to identify structures when the LOS
velocities become too small.

The second panel in Figure~\ref{fig:histograma_downflows2} shows the
distribution of the mean horizontal velocity of the downflowing
patches. This parameter has been calculated using well-defined
structures for which the direction of motion is clear.
Positive (negative) values represent motions away from (to)
  the umbra. As can be seen, most patches move outward. The median
horizontal speed is 85~m~s$^{-1}$. Finally, the last two panels of the
figure show the ratios of maximum to initial mean LOS velocity
  and continuum intensity.  About 75\% of the patches develop larger
  redshifts and brighten during their evolution. The median ratios are
  22\% and 3\%, so the velocity increases more than the brightness.

\begin{figure}[t]
	\begin{center}
	\includegraphics[trim=1cm 4cm 2cm 0cm, clip=true,width=.4\textwidth]{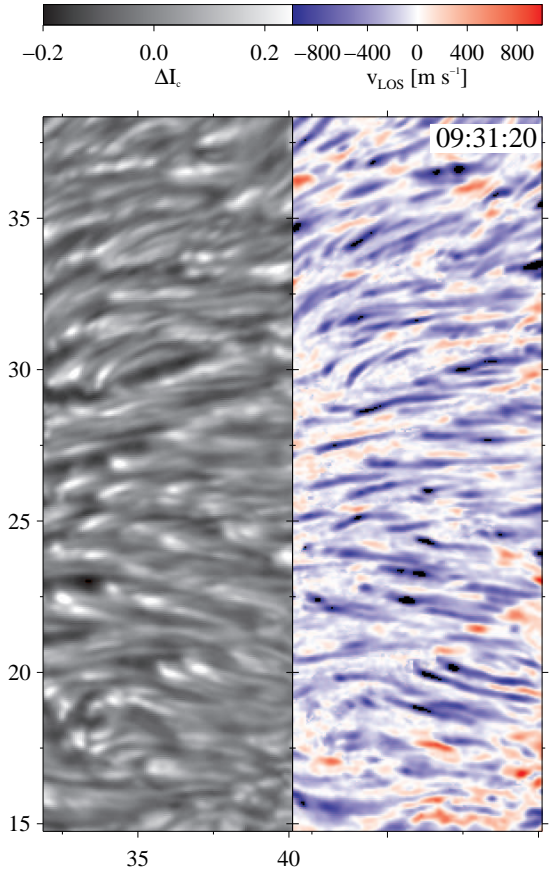}
	\caption{Continuum intensity fluctuations (left) and bisector
          velocity at the 70\% intensity level (right) in the
          center-side penumbra of AR 11302. The x- and y-axes indicate
          the position of the region in
          Figure~\ref{fig:continuo_dcarrow}.}
	\label{fig:fluctuaciones}
	\end{center}
\end{figure}

To study the relation between brightness and LOS velocity we performed
an analysis of local fluctuations by subtracting smoothed versions of
the continuum intensity filtergrams from the filtergrams themselves. The
smoothing was done over 30 pixels, i.e., 1\farcs8. The left panel 
of Figure~\ref{fig:fluctuaciones} shows the resulting intensity
fluctuations in the region of the center-side penumbra analyzed
before. Penumbral filaments are clearly seen as bright structures
standing above the mean continuum intensity by some $0.1 I_{\rm QS}$.
Penumbral grains at the head of the filaments show intensity
enhancements that can exceed $0.3 I_{\rm QS}$. Between the filaments
there are dark lanes with negative variations (around $-0.1 I_{\rm
  QS}$).

The right panel of Figure~\ref{fig:fluctuaciones} displays the
corresponding bisector velocities at the 70\% intensity level. As can
be seen, the blueshifted flow channels are surrounded by regions where
the velocities are more shifted to the red. The strongest redshifts
however do not always coincide with the darkest areas, as they often
occur close to the edges of the filaments.

\begin{figure}[t]
	\begin{center}
	\includegraphics[bb=20 0 445 382,width
	=.45\textwidth]{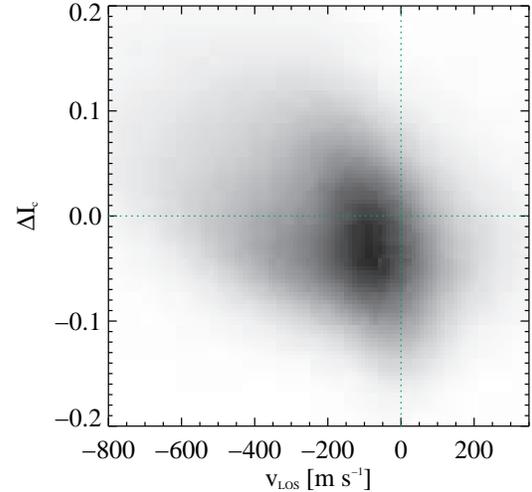}
	\caption{Scatter plot of continuum intensity fluctuations vs LOS velocity in the center-side 
penumbral region displayed in Figure~\ref{fig:fluctuaciones}. The full time sequence is used here.} 
	\label{fig:densidad}
	\end{center}
\end{figure}

Figure~\ref{fig:densidad} shows a scatter plot of continuum intensity
fluctuations vs LOS velocity for the region displayed in
Figure~\ref{fig:fluctuaciones}. This plot quantifies the behavior
described above: pixels with redshifts tend to exhibit negative
intensity fluctuations, and vice versa. In fact, the relation is
more or less linear, suggesting a correlation between temperatures
and flows: the hotter (brighter) structures harbor the largest upflows
(blueshifts), whereas the colder structures tend to be ones with the
stronger downward velocities (redshifts).  This is just the behavior
expected from convective flows and has been observed also by
\cite{2007ApJ...658.1357S} and \cite{2011Sci...333..316S}.

\section{Discussion}
\label{sec:discussion}

\subsection{Why have lateral downflows escaped detection so far?}

Sunspot penumbrae have been subject to intense scrutiny for years,
using all observational techniques, from monochromatic imaging to
Stokes spectropolarimetry, at high and moderate
spatial resolution, and high and low cadence. Yet, the lateral
downflows studied in this paper have consistently eluded detection
\citep[e.g.,][]{2007PASJ...59S.601J, 2008A&A...481L..13B, 2009A&A...508.1453F,
  2010ApJ...725...11B, 2011PhDT.......137F}.
Having established their properties, we can now provide explanations
why their identification was so challenging.

The first problem is our inability to determine the three components
of the velocity vector. Only the LOS component is accessible via
spectroscopic measurements. Thus, in order to detect vertical motions
unambiguously, it is necessary to observe sunspots as close as possible
to the disk center, so that the LOS coincides with the normal to the
local surface \citep{2011ApJ...739...35B}. However, this kind of
observations are very scarce.  Most measurements carried out so far
correspond to spots away from the disk center, where the velocity
field is dominated by the strong Evershed flow, effectively hiding
the weak lateral downflows.

Spectroscopy, and more so spectropolarimetry, is very demanding in
terms of exposure time. The long exposures that are required generally
mean lower spatial resolution because of stronger seeing degradation,
unless the observations are made from space. The highest resolution
measurements of sunspot penumbrae using a slit spectrograph were
performed by \citet{2010ApJ...725...11B} at the SST.  Exposure times
of 200~ms were needed and, despite the excellent seeing conditions,
the achieved spatial resolution was not better than some 0\farcs2.
This is barely sufficient to detect the largest downflow
patches, as the median width of these structures is only 0\farcs16.
Lateral downflows are also out of reach for the Hinode 
spectropolarimeter, because of its resolution of about 0\farcs32.
This may explain why \citet{2009A&A...508.1453F} did not detect them
in their careful analysis of sunspots at the disk center.

Of course, fingerprints of such weak downflows must be present in the
observed Stokes profiles even at moderate spatial resolution (e.g.,
the Doppler shifts they produce, their possible opposite polarities,
etc) but, being unresolved, separating them from the dominant signals
of the strong Evershed flows is not easy---not even for complex Stokes
inversions such as those performed by \cite{2003ASPC..307..301B},
\cite{2003A&A...410..695M},
\cite{2004A&A...427..319B}, \cite{2005A&A...436..333B}, or
\cite{2011A&A...525A.133B}. Assuming the penumbra to be a
  micro-structured magnetic atmosphere, however,
  \cite{2005ApJ...622.1292S} reported downflows and opposite magnetic
  polarities all across the penumbra from the inversion of
  spectropolarimetric measurements. These downflows were found to have
  velocities of up to 10~km~s$^{-1}$.
More recently, hints of the largest
downflows patches may have been obtained by \citet{2009A&A...508.1453F, 
2013A&A...550A..97F} using the Hinode spectropolarimeter.

The intermittent character of the lateral downflows, which often
appear only on one side of the filaments, and for short periods of
time (the median lifetime is 3.2 minutes), also imposes strong
limitations to slit instruments. The spectroscopic observations of
\citet{2010ApJ...725...11B}, for example, crossed several penumbral
filaments, but given the intermittency of the lateral downflows it was
unlikely that just a single slit position could catch many instances
of them, if any at all.  Thus, in retrospect, the chances of success
of those observations were low.

The key to detection is probably ultra-high spatial resolution. This
is why imaging spectropolarimetric observations provided the first
evidence of lateral downflows \citep{2011Sci...333..316S,
2011ApJ...734L..18J}. The measurements were subject to post-facto
image restoration to reach the diffraction limit of the SST, which at
$\sim$0\farcs13 made it possible to resolve the smallest velocity
structures ever. Still, corrections other than standard image
reconstruction had to be applied to bring out the lateral flows.  They
were meant to compensate for substantial amounts of stray light, and
in practice involved strong deconvolutions of the data. For many observers, such corrections lowered
the significance of the discovery.

Reaching ultra-high spatial resolution is not sufficient, however. 
One needs to subtract from the velocity measurements the oscillation
pattern known to affect sunspot penumbrae. This pattern consists of
large-scale velocity fluctuations with amplitudes of the order of
100~m~s$^{-1}$, superposed onto the actual penumbral flow field.  Not
removing the oscillations may bias the velocities to the point of
making some structures appear redshifted when in reality they do not
harbor downward motions. Conversely, genuine lateral downflows can be
effectively hidden by the oscillations during their blue excursions,
since the amplitude of the latter is not much smaller than the LOS
velocities of the former. 

\begin{figure}[t]
	\begin{center}
    	\includegraphics[trim = 2.8cm 7cm 2.2cm 3.8cm, clip=true, 
        width = 0.5\textwidth]{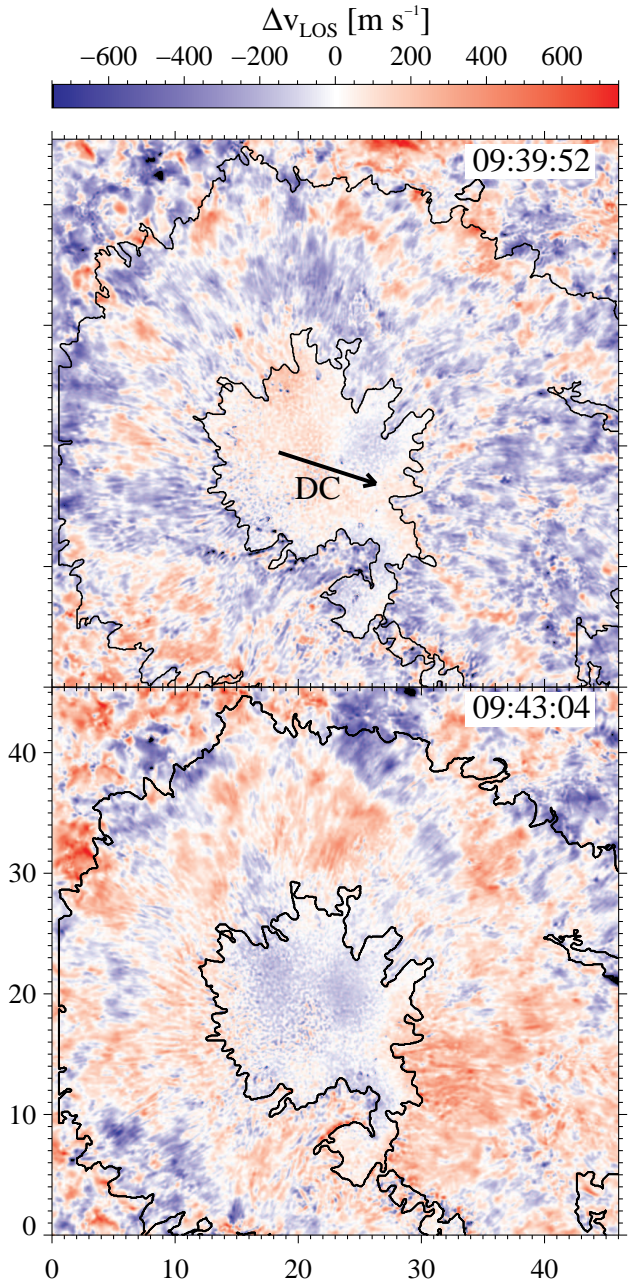}
        \caption{Amplitudes of the 5-minute oscillation in two
          frames of our Dopplergram time sequence, corresponding to
          phases in which the center-side penumbra is offset to the
          blue by $-92$~m~s$^{-1}$ (top) and to the red by
          88~m~s$^{-1}$ (bottom). The arrow marks the direction to the disk
          center.}
  		\label{fig:resta_osci_grande_paper}
  		\end{center}
\end{figure}

To subtract the oscillation pattern, one needs to secure time
sequences under stable seeing conditions. This is extremely
challenging.  All imaging spectroscopic measurements presented to date
are based on single snapshots and therefore they are affected by
oscillations in unknown ways, which casts doubts on the reliability of
the reported lateral downflows. To illustrate the importance of this
correction, Figure~\ref{fig:resta_osci_grande_paper} shows typical
velocity patterns created by oscillations in our sunspot. They
correspond to phases where the center-side penumbra is offset to the
blue by an average of $-92$~m~s$^{-1}$ (top panel) and to the
red by 88~m~s$^{-1}$ (bottom panel). These are the
offsets that could not be subtracted in previous investigations
because of the lack of temporal information.

The choice of spectral line is also relevant, although probably not of
prime importance. The discovery paper used the extremely weak
\ion{C}{1}~538~nm line \citep{2011Sci...333..316S}, which is thought
to be one of the deepest-forming lines in the visible but nearly
disappears in cool structures like the penumbral background that
surrounds the flow channels. In addition, \ion{C}{1} 538~nm is
affected by molecular blends and cannot be used to set up a reliable
velocity reference \citep{2012ASPC..463...99U}.
\cite{2011ApJ...734L..18J} employed the traditional \ion{Fe}{1} 630~nm
lines, which do not show these problems in the penumbra but have two
$O_2$ telluric lines distorting their red wings.  This is precisely
the most important region of the lines for observing downward motions
(redshifts), which again may raise doubts on the reliability of the
detection.  Other lines used in spectroscopic studies include
\ion{Fe}{1}~709.0~nm. This transition is narrow and therefore very
sensitive to velocities \citep{2005A&A...439..687C}, but a CN blend at
709.069~nm distorts its red wing. The \ion{Fe}{1} 617.3~nm line we
have observed is also very narrow, with the additional advantage of
its perfectly flat and clean red continuum. However, it seems to be
affected by an \ion{Eu}{2} 617.30~nm blend and two other molecular
lines in very cold umbrae \citep{2006SoPh..239...69N}.

On top of this all, we have found that a precise correction of subtle
instrumental effects is mandatory to unveil the existence of lateral
downflows, given the extremely weak Doppler signals they generate.
Having imaging spectropolarimetric observations at the diffraction
limit and excellent seeing for long periods of time does not help much
unless the etalon cavity errors are carefully corrected for.
Imperfect corrections produce an orange peel pattern
in the velocity maps, with amplitudes that easily exceed those of the
lateral downflows themselves. Here we have used the latest version of
the CRISP data reduction pipeline, where a new method is implemented
to remove the cavity errors. This method produces velocity maps of
unprecedented quality and remarkable smoothness.

In summary, the detection of lateral downflows is challenged by a
large number of factors that conspire together to hide them in most
but the best observations. A good knowledge of their properties, as
established here, will hopefully help design new strategies to further
study these features and their temporal evolution.

\begin{figure*}[t]
\begin{center}
    \includegraphics[trim = 6cm 4.5cm 7.35cm 0.5cm, clip = true, width
    = 1.\textwidth]{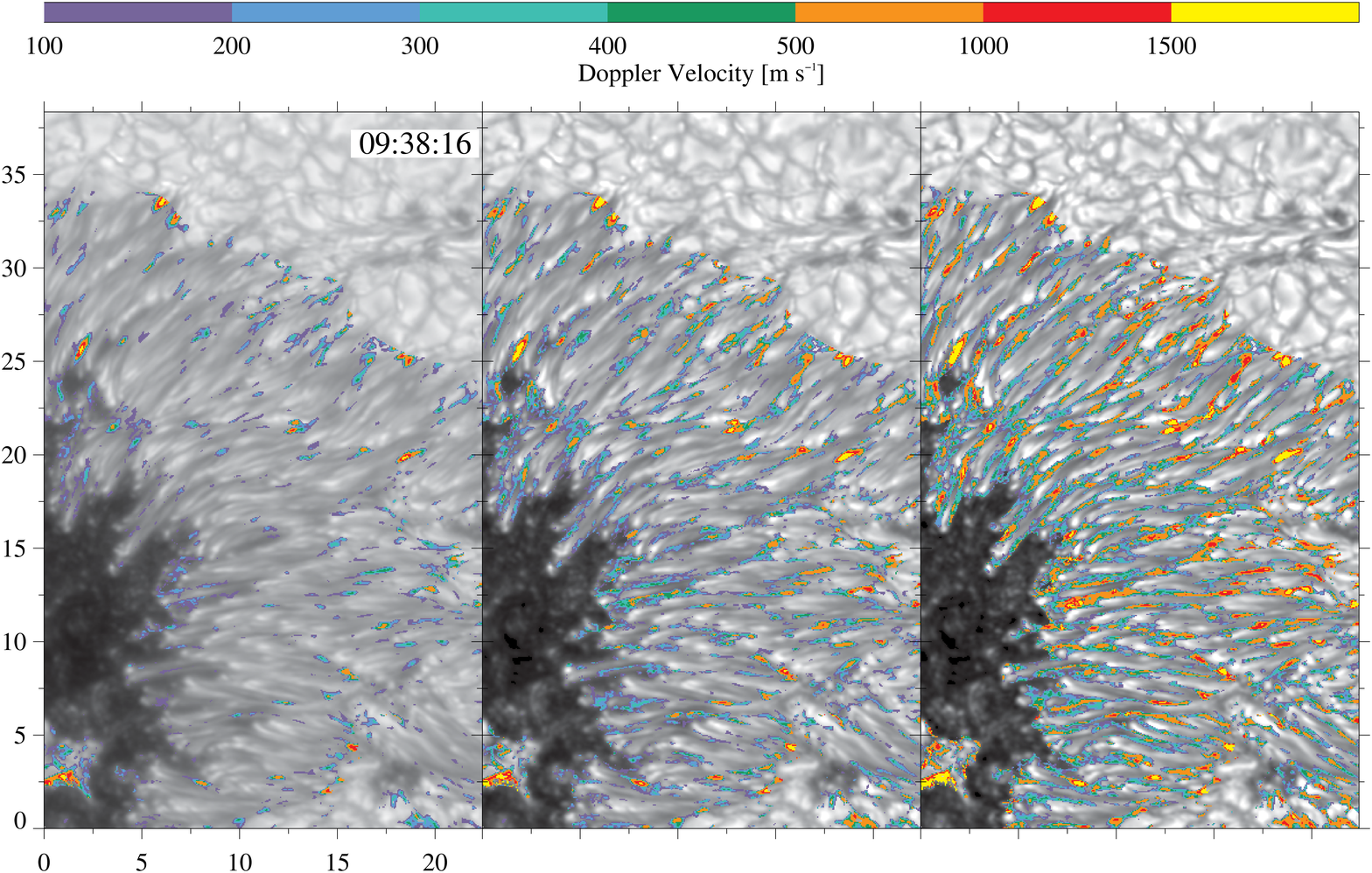}
    \caption{Comparison of the Doppler velocities to the red resulting
    from the original and deconvolved data.  From left to right:
    Original observations, compensation for 40\% stray light
    contamination, and compensation for 58\% contamination.}
\label{fig:rangos_velocidades}
\end{center}
\end{figure*}

\subsection{Stray-light Compensated Data}
\label{sub:straylight}

In this section we show the velocities resulting from our observations
after deconvolving them with different levels of stray light, to allow
direct comparisons with \cite{2011Sci...333..316S,
2013A&A...553A..63S} and \citet{2012A&A...540A..19S}.  The existence
of stray light contamination, its importance, and nature are still a
matter of debate (see \citealt{2012A&A...537A..80L}, \citealt{2013A&A...555A..84S}, 
and \citealt{2014A&A...561A..31S}).

We compensate for stray light in the same way as
\citet{2011Sci...333..316S}.  Assuming that $I_{\rm t}$ is the
true intensity emerging from a pixel at a given wavelength and
polarization state, the observed intensity $I_{\rm o}$ is
degraded by stray light as
\begin{equation}
\label{eq:deconvolution}
I_{\rm o} = (1 - \alpha) I_{\rm t} + \alpha I_{\rm t} \ast P,
\end{equation}
where $\alpha$ represents the fraction of stray light contamination
and $P$ is a Gaussian PSF with a full width at half maximum $W$. The
symbol $\ast$ denotes convolution. Therefore, to obtain the true
intensity one needs to deconvolve the original data with some values
of $\alpha$ and $W$.

Figure~\ref{fig:rangos_velocidades} shows the lateral redshifts
observed in the original Dopplergrams (left) and in the straylight
compensated Dopplergrams, for straylight contaminations of 40\%
(center) and 58\% (right), and FWHM values of $W= 1\farcs8$ and
$W=1\farcs2$, respectively. The velocities correspond to the bisector
shifts at the 70\% intensity level. They are overplotted on the
continuum images resulting from the different deconvolutions.

As expected, the deconvolved continuum images show higher contrasts
than the original data. The granulation pattern stands out more
prominently as the straylight fraction increases, primarily because
the intergranular lanes become darker. Also the penumbral filaments
and the outer penumbral border are more clearly defined. In the
deconvolved Dopplergrams, umbral dots, penumbral grains, and dark cores
of penumbral filaments show higher contrasts than in the original
data.

When no stray-light compensation is carried out (left panel in
Figure~\ref{fig:rangos_velocidades}), the lateral downflows detected
in the center-side penumbra have typical velocities of less than
300~m~s$^{-1}$. At the tail of penumbral filaments, we see patches
with velocities that often exceed 1500 m~s$^{-1}$ and correspond to
Evershed flows returning back to the solar surface. The velocities are
always stronger at the center of the patches, for both types of
structures.  This translates into concentric LOS velocity contours
(most easily seen in the case of the final downflows).

The straylight compensated data (center and right panels in
Figure~\ref{fig:rangos_velocidades}) show the same redshifted patches,
with larger areas and LOS velocities, plus new patches that were not
visible originally and show up only after deconvolution.  Correcting
the data for a straylight contamination of 40\% increases the LOS
velocities of the lateral downflows by a factor of two on average. 
\citet{2012A&A...540A..19S} found a similar behavior.
The downflows associated with returning Evershed flows undergo
slightly smaller enhancements. As expected, the velocity increase is
even stronger in the data compensated for a straylight contamination
of 58\%, with individual lateral downflows reaching up to 475~m~s$^{-1}$ 
compared to the $\sim$150~~m~s$^{-1}$ of the original data.

Our results suggest that the lateral downflows detected by
\citet{2011Sci...333..316S}, \citet{2012A&A...540A..19S} and
\citet{2013A&A...553A..63S} in deconvolved images are probably not
mathematical artifacts. Some of those redshifts are actually visible
in the original observations, and so they are not created by noise
amplification or by the data deconvolution \citep[see the
examples in][]{2013A&A...555A..84S}. We cannot say much about the
other structures that show up in the deconvolved images, as they are
the result of enhancing inconspicuous signals in the original
filtergrams.

\subsection{Interpretation}
\label{sub:interpretation}

The lateral downflows studied in this paper are compatible with
several magnetoconvection modes in sunspot penumbrae. We briefly 
describe them in the following.

\citet{1961ApJ...134..289D} suggested that penumbral
filaments represent convective rolls (cells) lying next to each other,
with their long dimension being parallel to the horizontal component
of the magnetic field. Two of these rolls would form a filament as
they turn in opposite directions, creating a bright upflow in the
center and two downflows on the external sides of the rolls. This
model was ruled out when confronted with observations
\citep[see][]{2011LRSP....8....3R}.  One of the main problems
was the lack of evidence for downflows at the filament edges.
Our analysis shows them unambiguously, which removes this drawback.

However, lateral downflows may also happen in penumbral flux tubes.
\citet{2007A&A...471..967B} derived the thermodynamic and magnetic
configuration of flux tubes in perfect mechanical equilibrium with the
surroundings. He found that flux tubes are stable structures if the
magnetic field is mainly aligned with the tube axis and shows a small
azimuthal component, i.e., if the field is slightly twisted. In the
plane perpendicular to the tube axis, the magnetic field topology
resembles two convective rolls through which gas can flow aligned with
the magnetic field. The flow would be mainly radial but would show an
upward component at the center of the tube and downflows on the sides.
Actually, this flow pattern is the one to be expected, since force
balance requires the central part of the tube be hotter than the
external part, which would naturally trigger convective motions as
described above.  The model by \citet{2007A&A...471..967B}
  produces penumbral filaments with central dark cores and is
  consistent with all sunspot magnetic field measurements 
  performed to date.

Finally, the lateral downflows may also be the result of overturning
convection in the penumbra, as proposed by 
  \citet{2008ApJ...677L.149S}.  According to this model, penumbral
filaments are elongated convective cells with two distinct velocity
components. Upflows emerge into the surface at the filament head and
progressively get more inclined along the filaments until they sink at
their tails.  At the same time, the upward flows turn over laterally,
becoming downdrafts at the filament edges.  The resulting convective
pattern is very similar to that observed in the quiet Sun, except for
the existence of a preferred direction ---the one defined by the
sunspot magnetic field.

From a pure kinematic point of view, our observations cannot rule out
any of these models. Additional criteria need to be invoked to discern
between the different scenarios (for example, the existence and
spatial distribution of opposite polarities). We note, however, that
the mechanism of overturning convection seems to be backed up by the
latest 3D numerical simulations
\citep[e.g.,][]{2009ApJ...691..640R,2011ApJ...729....5R,
  2012ApJ...750...62R}.  Thus, we tentatively give more credence to
that scenario, with the caveat that the lateral downflow velocities
derived from the data are significantly smaller than those in the
simulations ($\sim$210~m~s$^{-1}$ vs 1~km~s$^{-1}$). The mismatch
persists even when stray light is corrected for.

Independently of the driver of the lateral downflows, the three models
predict the existence of a flow connecting them with the central
upflows. Such an azimuthal flow (perpendicular to the filament axis)
might be responsible for the so-called twisting motions that are seen
as dark lanes moving diagonally from the center to the limb side of
penumbral filaments \citep{2007Sci...5856.318}. Several mechanisms
have been proposed to explain these twisting motions
\citep{2008A&A...488L..17Z, 2010ApJ...722L.194B, 2010A&A...521A..72S,
  2012ApJ...752..128B}.  Therefore, the next step to corroborate if
they are tracers of the azimuthal component of the velocity field is
to study them by spectroscopic means.  This analysis could settle the
problem of the gas flow in penumbral filaments at photospheric levels.

\section{SUMMARY AND CONCLUSIONS}
\label{sec:conclusions}

In this paper, we have analyzed the dynamics and temporal evolution of
a sunspot penumbra using high-resolution spectropolarimetric
measurements in the \ion{Fe}{1}~617.3~nm line. The excellent seeing
conditions, accurate data reduction, precise wavelength calibration,
high temporal cadence and the location of the sunspot very close to
the disk center all contribute to make this an unprecedented dataset
for sunspot studies.

We report the ubiquitous presence of lateral downflows in penumbral
filaments. For the first time, these elusive features are detected
without correcting the data for stray light, avoiding the
controversies associated with such methods
\citep{2013A&A...555A..84S}.  We have removed the undesired imprint of
p-modes from the Dopplergrams using Fourier filtering. The resulting
velocity maps are less veiled and signals are more stable, as p-modes
can make large areas appear completely blueshifted (hiding the lateral
downflows) or redshifted (producing artificial downflows).  This
correction has not been performed previously and it is
important.

Lateral downflows appear in our Dopplergrams as elongated redshifted
patches surrounding the upflowing penumbral filaments. In the
centerside penumbra, they stand out conspicuously over the dominant
blueshifts associated with the Evershed flow. By studying
examples in different parts of the spot we have been able to rule out
radial and azimuthal motions as the cause of the redshifts. We
summarize the properties of these structures as follows:
\begin{itemize}
\item The downflows are mostly located in dark areas next to penumbral
  filaments, but some of them occur on the bright filament edges.

\item The LOS velocities observed at the $70\%$ bisector level
  typically range from 150 to 300~m~s$^{-1}$. Occasionally, stronger
  downflows exceeding $500$~m~s$^{-1}$ are detected.
  
\item We find an exponentially decreasing distribution of lifetimes,
  with nearly all downflows lasting less than 6~minutes.  The median
  lifetime is 3.2~minutes.  Within this time span, they can fragment,
  disappear, and re-appear. They seem to follow the wiggle of the
  filaments they are associated with. We speculate that forces derived
  from horizontal pressure balance are somehow responsible for this
  behavior.
   \item Lateral downflows show a median length of 0\farcs53 and a
   width of 0\farcs16. Therefore, high spatial resolution is
   needed to detect them.
 \item In the inner and middle penumbra, the downflowing patches move
   outward with a median horizontal velocity of 85~m~s$^{-1}$, whereas
   bright penumbral grains move inward \citep[][and
   others]{1973SoPh...29...55M, 2001A&A...380..714S,
     2006ApJ...646..593R}. The outward horizontal motion decreases the
   LOS velocity observed in the lateral downflows, so the actual flow
   speeds could be larger than those indicated by Doppler
   measurements.
 \item The lateral downflows detected in the original Doppler maps are
   found at the same locations in straylight-compensated data, but
   with enhanced speeds. Thus, our results suggest that at least some
   of the downflows seen in data corrected for stray light are not
   mathematical artifacts.
\end{itemize}

We have compared the continuum intensity fluctuations and the
corresponding LOS velocities in the center-side penumbra. Similarly to
\citet{2007ApJ...658.1357S} and \citet{2011Sci...333..316S},
we find that blueshifted areas are generally bright and redshifted
areas tend to be darker. These two quantities seem to follow a linear
relation, consistent with the operation of some kind of
magnetoconvection in the penumbra.

As other authors, we observe upflows along penumbral filaments and
downflows at their tails. The upflows return to the solar surface also
at the edges of the filaments, producing lateral downdrafts.  This
supports the existence of overturning convection in the penumbra
\citep{2006A&A...460..605S,2008ApJ...677L.149S}. The detection of the
lateral downflows has been challenging because they are very weak and
small.  We suggest a scenario where the lateral downflows are produced
by elongated convection: adjacent convective cells (filaments)
have a predominantly outward flow that returns to the solar surface at
their tails, while overturning convection is found along the edges of
the filaments.  There, matter from neighboring cells is strongly
squeezed and allowed to fall down.  

\citet{2005ApJ...622.1292S} was the first to infer the
  existence of downflows all over the penumbra from the inversion of
  moderate resolution spectropolarimetric data.  However, it is not
  clear to us that those downflows and ours represent the same
  phenomenon, given their very different spatial scales (optically
  thin versus optically thick) and speeds (tens of km~s$^{-1}$ versus
  a few hundred m~s$^{-1}$).

The velocities we observe in the downflows are significantly smaller
than those resulting from the simulations, even after correcting the
data for stray light. For this reason, the association of lateral
downflows with overturning convection is still not completely
unambiguous. In fact, other theoretical scenarios, such as the
convective rolls proposed by \citet{1961ApJ...134..289D} or the
twisted horizontal magnetic flux tubes of \citet{2007A&A...471..967B},
are also compatible with our observations from a pure kinematical
point of view. Distinguishing between them requires a thorough 
analysis of the vector magnetic field in penumbral filaments and 
a comparison with the velocity field at similar spatial scales.

\acknowledgments

Financial support by the Spanish Ministerio de Econom\'{\i}a y
Competitividad through grants AYA2009-14105-C06-06,
AYA2012-39636-C06-05, and ESP2013-47349-C6-1-R, including a percentage
from European FEDER funds, is gratefully acknowledged.  This paper is
based on data acquired at the Swedish 1-m Solar Telescope, operated by
the Institute for Solar Physics of Stockholm University in the Spanish
Observatorio del Roque de los Muchachos of the Instituto de
Astrof\'{\i}sica de Canarias. This research has made use of NASA's
Astrophysical Data System.

\end{document}